\documentclass[%
 reprint,
amsmath,
amssymb,
aps,
]{revtex4-2}

\usepackage{graphicx}			
\usepackage{dcolumn}			
\usepackage{bm}					

\usepackage{color}
\usepackage{xcolor}

\begin{document}

\preprint{APS/123-QED}

\title{\bfseries Elastic $dcs$ of $ep$-scattering fitted via the $dcs$ \\ of $eq$-scatterings with cloud-covering effects}

\author{Jingle B. Magallanes}														
\altaffiliation{Also a researcher at  Premier Research Institute of Science and Mathematics (PRISM)}				
\email{emails: jingle.magallanes@g.msuiit.edu.ph, jbmagallanes@gmail.com}

\author{Jinky B. Bornales}															
\affiliation{Department of Physics, Mindanao State University - Iligan Institute of Technology, Iligan City 9200 Philippines}

\author{Ren\'e Luna-Garc\'\i{}a}														
\affiliation{Centro de Investigaci\'on en Computaci\'on, Instituto Polit\'ecnico Nacional, M\'exico City 07738 M\'exico}
\date{\today}																

\begin{abstract}
The angular-averaged differential cross section ($dcs$) of the elastic electron-proton ($ep$) scattering, covering $Q^2 < 1.0GeV^2$, was fitted via a combined modified $eq$-scatterings where $q$ is a point particle. The modifications represent the cloud-covering effects to $q$. An energy-decaying ratio ($edr$) was derived by inspecting the generated $dcs_{ep}$ from the form factor data gathered at Mainz Microtron (A1-Collaboration) and Continuous Electron Beam Accelerator Facility (Jefferson Laboratory) when compared to the $dcs_{eq}$ with modified relativistic recoil factor. The diminishing cloud layer, $edr$, has a decay rate of $-2.8$ for the data sets under investigation. The formulated SBM and SEM fitting models use the bare and effective $u$ and $d$-quark masses, respectively, while SCBM and SCEM integrate other considerations. Three comparison methods were used and all of them favor the models with other additional considerations. SCEM was the most favored model in general.
\end{abstract}


\maketitle	



\section{Introduction}
Electron-nucleon scattering has been used to extensively measure the nucleon's electromagnetic form factors to study the charge and magnetization distributions \cite{YeZ7778152018}. For this, it is important to measure the scattering's differential cross section ($dcs$) since it is proportional to the probability for any given reaction or process to occur. The objective of this study is to demonstrate a fitting model to the angular-averaged $dcs$ of the elastic electron-proton ($ep$) scattering, $dcs_{ep}$, generated from different form factor data sets covering the transfer momentum, $Q < 1GeV$. 

Initially, it was thought that fitting the $dcs_{ep}$ through electron-point particle ($eq$) scatterings would be impossible since the proton is definitely not a point particle as characterized by the form factors. However, it could and would be possible by putting some cloud-covering effects on the point particle $q$. Inasmuch as, at low-energy Quantum Chromodynamics (QCD) where both perturbation theory and asymptotic freedom are not possible, there are significant collective interactions between the valence and sea quarks; and the effects are in the form of cloud coverings. The valence quarks get surrounded by some dense concentration of virtual quarks and gluons.  When probed at low energy, this cloud is the high energy barrier to the core of the proton.

For the range of transfer momenta in consideration, $eq$-scattering would have to be masked by modifications mantling the particle. This includes the modifications in $dcs_{eq}$'s recoil factor (fixed cloud layer) and the energy dependent ratio (diminishing cloud layer) between $dcs_{ep}$ and $dcs_{eq}$.

\section{The Electron-Proton ($ep$) Scattering}

The elastic $ep$-scattering is one of the fundamental interactions used in the understanding of the structure and the build-up of hadronic physics \cite{Dainton55781999}. It is called Mott or no-structure ($ns$) scattering when it is the electron that is scattered by the point-particle nucleus. Electrons are very light; with high energies, they can penetrate further into the nucleus. However, they couple to the nuclear magnetic field because they have nonzero spin, an effect carried by the final term in the $dcs$ given in Equation \ref{NSMottDCS}. This equation also contains the ratio between the final ($E'$) and initial ($E$) energies of the electron called the relativistic recoil factor of the nucleus. The cross-section is denoted by $\sigma_{ns}$ for the Mott scattering:
\begin{eqnarray}
{\frac{d \sigma}{d \Omega}}_{Mott} && = \sigma_{ns} \nonumber \\
&& = \frac{(Z_1 Z_2 \alpha)^2}{4k^2 sin^4 \left( \frac{\theta}{2} \right)} \left( \frac{E'}{E} \right) \left\lbrace 1 - v^2 sin^2 \left( \frac{\theta}{2} \right) \right\rbrace
\label{NSMottDCS}
\end{eqnarray}
where
\begin{equation}
\frac{E'}{E} = \frac{1}{1 + \frac{2E}{M} sin^2 \frac{\theta}{2}}
\label{EquationRecoilFactor}
\end{equation}
and
\begin{equation}
E - E' = \frac{-\overline{q}^2}{2M} = \frac{Q^2}{2M}
\label{EquationEnergyDifference}
\end{equation}
with $M$ being the mass of the nucleon. The electron has to release a virtual photon as a necessary condition in order to probe the proton with an energy equal to the difference between the electron's initial and final energies, given by Equation \ref{EquationEnergyDifference}, where $\overline{q}^2$ is the square of the transfer momentum and $-\overline{q}^2 = Q^2$.

Electron scattering has been deeply studied over the years and there are two cases: the elastic scattering characterized by the electromagnetic form factors and the deep inelastic scattering characterized by the structure functions. Electromagnetic form factors of the proton provide some of the first information of its size and distribution of charge and magnetization. Moreover, the observation of unexpected behavior in form factors and structure functions has also brought new understanding of the strong interaction.

The electron being a point-particle has the simple vertex, $\gamma_\mu$, and its current takes the form $j_\mu = -e \bar{u} (k') \gamma_\mu u (k)$
while the proton has a vertex, $\Gamma^\mu$, with a current expressed using form factors parameterizing its internal structure. Also, the proton current must be a Lorentz-invariant four-vector that satisfies the parity and current conservation of the electromagnetic interaction. Hence, for a single-photon exchange, two form factors are allowed in the vertex and the current is given by
\begin{eqnarray}
J^\mu && = e \bar{v} (p') \Gamma^\mu v (p) \nonumber \\ 
&& = e \bar{v} (p') \left[ F_1(q^2) \gamma^\mu + \frac{i \kappa}{2M} F_2(q^2) \sigma^{\mu\nu} q_\nu \right] v (p)
\end{eqnarray}
where $F_1(q^2)$ is the Dirac form factor corresponding to the helicity-conserving current; $F_2(q^2)$ is the Pauli form factor corresponding to the helicity-flip current; $\kappa = 1.793\mu_N$ is the proton anomalous magnetic moment; $M$ is the proton nucleon mass; and $\sigma_{\mu\nu} = 2i\left[ \gamma_\mu, \gamma_\nu \right]$. For $q^2\rightarrow 0$, $F_1(0) = F_2(0) = 1.0$ in the non-relativistic limit and the proton is treated as a point-particle where the virtual photon is insensitive to the proton's internal structure.

The $dcs$ becomes
\begin{eqnarray}
\frac{d \sigma}{d \Omega} & = & \frac{\vert j_\mu \frac{1}{q^2} J_\mu \vert^2}{4\left( (k \dot{p})^2 - m^2 M^2 \right)} (2 \pi)^4 \delta^4(k' - k + p - p') \nonumber \\
 && \times \frac{d^3k' d^3p'}{(2 \pi)^3 2E' (2 \pi)^3 2 (M+\omega)}. 
\end{eqnarray}
where the conservation of momentum is assured by the delta functions. Integrating over the relevant variables; averaging initial spin states; and summing over final ones, the $dcs$ as a function of the scattering angle $\theta$ becomes
\begin{eqnarray}
&& \frac{d \sigma}{d \Omega} = \frac{(Z_1 Z_2 \alpha)^2}{4k^2 sin^4 \left( \frac{\theta}{2} \right)} \left( \frac{E'}{E} \right) cos^2 \frac{\theta}{2} \nonumber \\
&& \times \left[ \left( F_1^2 + \frac{\kappa^2 Q^2}{4M^2} F_2^2 \right) + \frac{Q^2}{2M^2} \left( F_1 + \kappa F_2 \right)^2 tan^2 \frac{\theta}{2} \right]. 
\label{DCSFormFactorsF1F2}
\end{eqnarray}
This can simplify to the structureless Mott cross section multiplied with the form factor term where $(1-v^2 sin^2 \frac{\theta}{2}) \rightarrow cos^2 \frac{\theta}{2}$, for relativistic electrons. If the proton were a point charge, its $dcs$ would have only been
\begin{equation}
\frac{d \sigma}{d \Omega} = \frac{(Z_1 Z_2 \alpha)^2}{4k^2 sin^4 \left( \frac{\theta}{2} \right)} \left( \frac{E'}{E} \right) \left[ cos^2 \frac{\theta}{2} + \frac{Q^2}{2M^2} sin^2 \frac{\theta}{2} \right].
\label{DCSEPScatteringPoint}
\end{equation}
To avoid the interference between $F_1$ and $F_2$ in Equation \ref{DCSFormFactorsF1F2}, the structure-dependent part of the cross section can be rewritten in terms of the electric and magnetic form factors $G_E(Q^2)$ and $G_M(Q^2)$ \cite{SachsRG12622561962} where $G_E(Q^2) = F_1(Q^2) - \kappa \tau F_2(Q^2)$
and $G_M(Q^2) = F_1(Q^2) + \kappa F_2(Q^2)$.
Then, with $\tau=Q^2/4M^2$, the $dcs$ becomes
\begin{eqnarray}
\frac{d \sigma}{d \Omega} = && \frac{(Z_1 Z_2 \alpha)^2}{4k^2 sin^4 \left( \frac{\theta}{2} \right)} \left( \frac{E'}{E} \right) cos^2 \frac{\theta}{2} \nonumber \\
&& \times \left[ \frac{G_E^2 + \tau G_M^2}{1+\tau} + 2 \tau G_M^2 tan^2 \frac{\theta}{2} \right]
\label{DCSEPScatteringFinite} 
\end{eqnarray}
which can be further simplified to
\begin{equation}
\frac{d \sigma}{d \Omega} = \sigma_{ns} \frac{1}{1+\tau}\left[ G_E^2 + \frac{\tau}{\epsilon}G_M^2 \right]
\end{equation}
where
\begin{eqnarray}
1/\epsilon & = & [1+2(1+Q^2/4M^2)tan^2(\theta/2)] \nonumber \\
& = &[1+2(1+\tau)tan^2(\theta/2)];
\label{NoneForNow} 
\end{eqnarray}
$\epsilon$ is an angular variable. In the non-relativistic limit, $Q \rightarrow 0$, these form factors are just the Fourier transforms of the charge and magnetization distributions \cite{HalzenFMartinAD1984},
\begin{equation}
F_{nr}(Q^2) = \int \rho(\overrightarrow{r}) e^{-\overrightarrow{Q} \cdot \overrightarrow{r}} d^3 \overrightarrow{r}.
\end{equation}
Dipole form factor,
\begin{equation}
G_D(Q^2) = \frac{1}{(1+a^2 Q^2)^{2}}
\end{equation}
comes out if the charge distribution is exponential, $\rho(r)= \rho_0 e^{-r/a}$, where $a$ is the scale of the proton radius given by $a^2=(0.71GeV^2)^{-1} $. If the charge and magnetic moment distributions are the same, then their transforms will be as well; and generally, the form factor ratio will be
\begin{equation}
\frac{\mu G_E (Q^2)}{G_M (Q^2)} = 1.0,
\end{equation}
which is known as the form factor scaling. For low $Q^2$, at which the electric and magnetic root mean square (rms) radii can be determined \cite{Bernauer1011031052420012010}, the form factors can be expanded as
\begin{equation}
\frac{G (Q^2)}{G (0)} = 1 - \frac{1}{6} \left< r^2 \right> Q^2 + \frac{1}{120} \left< r^4 \right> Q^4 - ... \: .
\end{equation}
The (rms) radius can be determined from the slope of the form factors at $Q^2=0$ with
\begin{equation}
\left< r^2 \right> = - \frac{6}{G (0)} \frac{d G (Q^2)}{d Q^2}|_{Q^2=0}.
\end{equation}

\section{Low energy form factor data}

Form factors can be extracted via Rosenbluth Extraction Method \cite{RosenbluthMN7941950, BergerC35871971,JohnsonMJ2013}. The form factor ratio $\frac{\mu G_E}{G_M} $ is $\sim 1.0$ at lower energies with the world data in \cite{BergerC35871971,AndivahisL5054911994,WalkerRC4956711994,JanssensT1429221966,LittJ31401970} and this is consistent with the form factor scaling. Other methods of extractions are Polarization Transfer Method \cite{JonesM84139814022000} and Super-Rosenbluth Method \cite{QattanIA941423012005}. Previous Rosenbluth data used the Bosted global fit \cite{Bosted514091995} valid at $0<Q^2<7GeV^2$. Recently, the Global Fitting Procedure \cite{YeZ7778152018,LeeG920130132015} was used for the world data valid for $Q^2$ up to $\sim 30GeV^2$. There was already an attempt in separating the quark flavor contributions to the elastic form factors at low-energy, detailed in  \cite{CatesGD1062520032011}.

The  low momentum transfer data presented in \cite{Bernauer2012381012006, Bernauer000412214035042010} were determined from the measurements at the Mainz Microtron (MAMI) using the 3-spectrometer-facility of the A1-Collaboration taken in three periods between 2006 and 2007 using beam energies of $180$, $315$, $450$, $585$, $720$ and $855MeV$. The experiment covers $0.004GeV^2 < Q^2 < 1.0GeV^2$ with counting rate uncertainties below $0.2\%$ for most of the data points \cite{Bernauer1011031052420012010}. They separate the form factors by fitting a wide selection of models directly to the measured cross sections. Extensive simulations were done to test the validity of this method. Standard Rosenbluth extraction technique was used in comparing the results. Form factors determined via Rosenbluth Separation Method, Friedrich-Walcher Model, Polynomial Model and  Spline Model were used in this study. The details pertaining to the measurements and analyses can be found in \cite{Bernauer000412214035042010}.

For the experiment presented in \cite{JohnsonMJ2013}, high-precision proton Rosenbluth extractions using beam energies from $849MeV$ to $5.157GeV$ were performed covering a large range of transfer momenta, $0.40GeV^2< Q^2 < 5.76GeV^2$, focusing on the extremes of $\epsilon$ where two-photon exchanges (TPEs) occur. The experiment has higher momentum transfers than proton Rosenbluth experiments before this and provided higher precision at low momentum transfer. To reconcile the discrepancy of results with that of Polarization data, considerations were taken including the missing corrections from TPE, which are difficult to calculate, and the results from other experiments but are not expected to be valid at low $Q^2$. But for this study, only $Q^2 < 1GeV^2$ were considered and in which case TPE rarely happens, hence, correction will be not as reliable. For the purposes of comparing the models with the available data, only some of the Rosenbluth extracted values were included. The details pertaining to the experiment and data analyses are found in \cite{JohnsonMJ2013}.


\section{Implementations}

The averaged multiple-angle $dcs_{ep}$ was fitted by the modified $dcs$ of $eq$-scatterings at transfer momenta less than $1GeV$ where $q$ is a point particle. Since the proton is a finite particle, cloud-covering effects have to be carried-out on $q$. This also warrants that $m_{q} < m_{p}$, in terms of particle masses.  The quark flavor composition of the proton ($uud$) was the basis in the choice of masses for the $q$'s in the fitting models; taking the quark masses and their corresponding fractional charges. Accordingly, effective (low energy) quark masses \cite{YaoWM3312006,Griffiths2008} are assigned to $q$ for the transfer momentum in consideration, but it could also be assigned bare quark masses \cite{YaoWM3312006,PatrignaniC40101000012016} since the cloud-effect is already represented by the modifications. The relativistic recoil factor of the angle and spin averaged $dcs$ of $eq$-scattering was modified using the proton mass as a parameter. Overlapping of the electron wave functions, spin-spin interactions, and color interactions were also considered in coming-up with the fitting models but arbitrarily not quantitative yet. The form factors derived from experiments at Mainz Microtron (MAMI) \cite{Bernauer2012381012006, Bernauer000412214035042010} and Continuous Electron Beam Accelerator Facility (CEBAF, JLab) \cite{JohnsonMJ2013} were used to generate the data for $dcs_{ep}$.

The angular-averaged $dcs_{ep}$ were generated via Equation \ref{DCSEPScatteringFinite} in ROOT Data Analysis Framework \cite{BrunRademakers9638981861997} platform. Raw $dcs$ of $eq$-scattering with $q$ having the mass of $u$-quark ($dcs_{eu}$) and $d$-quark ($dcs_{ed}$) were also simultaneously generated using the same random numbers via Equation \ref{DCSEPScatteringPoint}. A total of $2000$ data points each for $dcs_{ep}$, $dcs_{eu}$ and $dcs_{ed}$ were gathered at random various scattering angles from $0^o$ to $180^o$ for each corresponding particular transfer momentum in the experimental data considered. The energy-decaying ratios, which decreases as photon energy increases, between $dcs_{ep}$ and $dcs_{eq}$ were then determined and incorporated back to the $dcs_{eq}$ modifying them further. New data points were generated and then re-analyzed.

Equation \ref{EquationRecoilFactor} is the relativistic recoil factor and this is due to the recoil of the target particle during the interaction \cite{HalzenFMartinAD1984,Martin22012}. Its modification has a significant change to the $dcs$, acting like a fixed layer of cloud, as it shifts the $dcs_{eq}$ distribution vertically and closer to the $dcs_{ep}$ when the mass used is similar to that of proton. At a particular $Q^2$ and considering an angle-averaged $dcs$, the recoil factor is a constant. This materializes the proton mass as a parameter to the fitting model.

Correspondingly, the averaged points in the same transfer momentum for $dcs_{ep}$ and $dcs_{eq}$ were compared. There is a decreasing ratio, between $dcs_{ep}$ and $dcs_{eu}$ and more so with $dcs_{ed}$, behaving exponentially. A dimensionless energy-decaying ratio ($edr$) of the form $A e^{-r Q^2}$ was found for the investigated Rosenbluth form factor data sets with $A$ as the amplitude and $r$ as the decay rate, see TABLE \ref{EnergyDecayingRatio}. There are differences in the amplitudes of the $edr_{d**}$ and $edr_{u**}$ but the decay rate for each data set is the same. It should be noted that the data set from \cite{Bernauer000412214035042010} has 27 selected data points while \cite{JohnsonMJ2013} only has 6 data points; experiments from which the data sets were taken have different considerations.

Combined fitting models with contributions from both $dcs_{eu}$ and $dcs_{ed}$ underpins the quark flavor composition of the proton. Additionally, the weight of the modified $dcs_{eu}$ and $dcs_{ed}$ contributions can be affected by the overlapping of the electron's initial and final wave functions, spin-spin interactions of the electron and proton, and color interactions of the quarks inside the proton, and other considerations. For instance, the contributions can be arbitrarily set to be $80\%$ instead of 2/3 for $dcs_{eu}$  and $20\%$ instead of 1/3 for $dcs_{ed}$.

\begin{table}[htbp]
\caption{\label{EnergyDecayingRatio} Energy-Decaying Ratio: The $edr$ was derived from the comparison of the data gathered by the known method of extracting form factors at low transfer momentum (Rosenbluth Extraction Method) to the $dcs_{eu}$ and $dcs_{ed}$ at fixed transfer momentum.} 
\begin{ruledtabular}
\begin{tabular}{ccccc}
\begin{tabular}[c]{@{}l@{}} \textbf{Form Factor} \\ \textbf{Data Sets} \end{tabular} & \textbf{$A$} & \textbf{$r$} & \textbf{Form} & \textbf{Notation} \\
\hline
\begin{tabular}[c]{@{}l@{}} Rosenbluth \\ Separation \\ Data \cite{Bernauer000412214035042010} \\ $ep$-$eu$ with \\ bare mass \end{tabular} & 3.50 & 2.8 & $3.50 e^{-2.8 Q^2}$ & $edr_{ubs}$ \\
\hline
\begin{tabular}[c]{@{}l@{}} Rosenbluth \\ Separation \\ Data \cite{Bernauer000412214035042010} \\ $ep$-$ed$ with \\ bare mass \end{tabular} & 14.0 & 2.8 & $14.0 e^{-2.8 Q^2}$ & $edr_{dbs}$ \\
\hline
\begin{tabular}[c]{@{}l@{}} Rosenbluth \\ Separation \\ Data \cite{Bernauer000412214035042010} \\ $ep$-$eu$ with \\ effective mass \end{tabular} & 2.40 & 2.8 & $2.40 e^{-2.8 Q^2}$ & $edr_{ues}$ \\
\hline
\begin{tabular}[c]{@{}l@{}} Rosenbluth \\ Separation \\ Data \cite{Bernauer000412214035042010} \\ $ep$-$ed$ with \\ effective mass \end{tabular} & 9.60 & 2.8 & $9.60 e^{-2.8 Q^2}$ & $edr_{des}$ \\
\hline
\begin{tabular}[c]{@{}l@{}} Rosenbluth \\ Extraction \\ Data \cite{JohnsonMJ2013} \\ $ep$-$eu$ with \\ bare mass \end{tabular} & 1.85 & 1.8 & $1.85 e^{-1.8 Q^2}$ & $edr_{ube}$ \\
\hline
\begin{tabular}[c]{@{}l@{}} Rosenbluth \\ Extraction \\ Data \cite{JohnsonMJ2013} \\ $ep$-$ed$ with \\ bare mass \end{tabular} & 7.40 & 1.8 & $7.40 e^{-1.8 Q^2}$ & $edr_{dbe}$ \\
\hline
\begin{tabular}[c]{@{}l@{}} Rosenbluth \\ Extraction \\ Data \cite{JohnsonMJ2013} \\ $ep$-$eu$ with \\ effective mass \end{tabular} & 1.45 & 1.8 & $1.45 e^{-1.8 Q^2}$ & $edr_{uee}$ \\
\hline
\begin{tabular}[c]{@{}l@{}} Rosenbluth \\ Extraction \\ Data \cite{JohnsonMJ2013} \\ $ep$-$ed$ with \\ effective mass \end{tabular} & 5.80 & 1.8 & $5.80 e^{-1.8 Q^2}$ & $edr_{dee}$ \\
\end{tabular}
\end{ruledtabular}
\end{table}

\section{Results}

When probed with very low energy, most if not all, hadrons are just point particles. Gradual increase in the probe energy reveals that they are actually extended particles. At low energy, the valence quarks are cloud covered constituent quarks and the proton would be a lump of clouds with an extended size. And, it is difficult to describe this lump without increasing the energy of the photon probe. The cloud, however, can be treated as an energy barrier through the core of the proton which can be diminished by increasing the energy probe.

TABLE \ref{EnergyDecayingRatio} tabulates the $edr$ for the Rosenbluth data sets \cite{JohnsonMJ2013,Bernauer000412214035042010}.
The amplitudes of the $edr$ were derived by separately comparing $dcs_{eu}$ and $dcs_{ed}$ to $dcs_{ep}$.
Compromising results of point to point comparison, corresponding to different transfer momenta, led to a concensus amplitude ratio of $\sim 4$.
One of the critical reasons being looked into is that, at very low transfer momenta, the ratio between $dcs_{eu}$ and $dcs_{ed}$ is predominantly affected by the ratio of the squares of their respective charges.
Thus, in order to close-in with $dcs_{ep}$, $dcs_{ed}$ have to be intensified by about four times as much as $dcs_{eu}$.
However, the transfer momentum, as it increments, also eventually affects the $dcs$ ratio in addition to the effects contributed by the assigned masses to the point particles; this aspect is open for more investigations.
Moreover, the amplitudes for $edr_{*e*}$ are lesser than $edr_{*b*}$ since, at the range of transfer momenta in consideration, the particles with effective masses are presumably having thinner clouds than those carrying their bare masses.
The decay rate of the diminishing cloud effect layer, $edr$, for each data set is constant.
It can be seen, however, that the decay rate for Rosenbluth form factor in \cite{Bernauer000412214035042010} is greater than in \cite{JohnsonMJ2013}.
The reason for this is speculated to be caused by either or both the experimental set-up considerations and of the statistical data size.

The $dcs_{ep}$ generated from the investigated Rosenbluth form factor data sets are compared to the  $edr_{u**} dcs_{eu}$ and $edr_{d**} dcs_{ed}$ and three ways of comparison were done|Ratio Test (averaging the ratios between the corresponding generated experimental data and fitting data) in TABLE \ref{RatioTestUD}, Absolute Difference (averaging the absolute differences between the corresponding generated experimental data and fitting data) in TABLE \ref{AbsoluteDifferenceUD} and Chi Test (square-root of the average of the squares of the differences between the corresponding generated experimental data and fitting data) in TABLE \ref{ChiTestUD}. Other form factor data sets were also used for comparison such as those determined by Friedrich-Walcher, Polynomial and Spline models with $68.3\%$ confidence level. The description of the other form factor models and the parameters for their best fits are found in chapter 7 and appendix J of \cite{Bernauer000412214035042010}.


For the Ratio Test in TABLE \ref{RatioTestUD}, the $dcs_{ep}$ generated from form factors of the models from \cite{Bernauer000412214035042010} were closer to the modified $dcs_{eu}$, except for the Rosenbluth Extraction Data of \cite{JohnsonMJ2013}, than to the modified $dcs_{ed}$ where the $q$'s assume bare masses. As expected, the $dcs_{ep}$ generated from Rosenbluth extraction method are the ones closer to $edr_{u**} dcs_{eu}$ and $edr_{d**} dcs_{ed}$ compared to the ones generated from other data sets. However, corresponding numbers as seen in TABLE \ref{RatioTestUD} are not in agreement among themselves which could be attributed to the differences in the experimental set-ups from which the two data sets were taken. The data from  \cite{Bernauer000412214035042010} were derived from the set-up that was intended for measurements using low beam energies while from \cite{JohnsonMJ2013} were measured from the set-up intended for higher beam energies.

\begin{table}[htbp]
\caption{\label{RatioTestUD} Ratio Test: The average ratio between the $dcs_{ep}$ generated from the different data sets to their corresponding $dcs_{eu}$ and $dcs_{ed}$ with $edr$ where bare quark masses (BM) are used and, separately, for effective quark masses (EM).} 
\begin{ruledtabular}
\begin{tabular}{ccccc}
\begin{tabular}[c]{@{}l@{}} \textbf{Form Factor} \\ \textbf{Data Sets} \end{tabular} & \begin{tabular}[c]{@{}l@{}} $ep$-$eu$ \\ \textbf{(BM)} \end{tabular} & \begin{tabular}[c]{@{}l@{}} $ep$-$ed$ \\ \textbf{(BM)} \end{tabular} & \begin{tabular}[c]{@{}l@{}} $ep$-$eu$ \\ \textbf{(EM)} \end{tabular} & \begin{tabular}[c]{@{}l@{}} $ep$-$ed$ \\ \textbf{(EM)} \end{tabular} \\
\hline
\begin{tabular}[c]{@{}l@{}} Rosenbluth \\ Extraction \cite{JohnsonMJ2013} \end{tabular} & 0.96964 & 0.96971 & 1.0003 & 0.99732 \\
\hline
\begin{tabular}[c]{@{}l@{}} Rosenbluth \\ Separation \cite{Bernauer000412214035042010} \end{tabular} & 1.0167 & 1.0173 & 0.98951 & 0.98742 \\
\hline
\begin{tabular}[c]{@{}l@{}} Friedrich-Walcher \\ Model \cite{Bernauer000412214035042010} \end{tabular} & 1.0504 & 1.0507 & 1.1520 & 1.1490 \\
\hline
\begin{tabular}[c]{@{}l@{}} Polynomial \\ Model \cite{Bernauer000412214035042010} \end{tabular} & 1.2053 & 1.2056 & 1.3455 & 1.3420 \\
\hline
\begin{tabular}[c]{@{}l@{}} Spline \\ Model \cite{Bernauer000412214035042010} \end{tabular} & 1.2134 & 1.2138 & 1.3558 & 1.3521 \\
\end{tabular}
\end{ruledtabular}
\end{table}

For the Absolute Difference in Table \ref{AbsoluteDifferenceUD}, the $dcs_{ep}$ generated from different data sets were more in agreement with $edr_{*e*} dcs_{e*}$ than $edr_{*b*} dcs_{e*}$ since the differences are much smaller in favor of the $dcs$ where quarks are assuming the effective masses. It can also be seen that all the $dcs_{ep}$ are in more agreement with $edr_{u**} dcs_{eu}$ than with $edr_{d**} dcs_{ed}$ except for the Rosenbluth Extraction Data \cite{JohnsonMJ2013}. Among the data sets from \cite{Bernauer000412214035042010}, the generated $dcs_{ep}$ from Friedrich-Walcher has the lowest average absolute difference.

\begin{table}[htbp]
\caption{\label{AbsoluteDifferenceUD} Absolute Difference: The average absolute difference between the $dcs_{ep}$ generated from the different data sets and their corresponding $dcs_{eu}$ and $dcs_{ed}$ with $edr$ where bare quark masses (BM) are used and, separately, for the effective quark masses (EM).} 
\begin{ruledtabular}
\begin{tabular}{ccccc}
\begin{tabular}[c]{@{}l@{}} \textbf{Form Factor} \\ \textbf{Data Sets} \end{tabular} & \begin{tabular}[c]{@{}l@{}} $ep$-$eu$ \\ \textbf{(BM)} \\ $\times 10^{-6}$ \end{tabular} & \begin{tabular}[c]{@{}l@{}} $ep$-$ed$ \\ \textbf{(BM)} \\ $\times 10^{-6}$ \end{tabular} & \begin{tabular}[c]{@{}l@{}} $ep$-$eu$ \\ \textbf{(EM)} \\ $\times 10^{-6}$ \end{tabular} & \begin{tabular}[c]{@{}l@{}} $ep$-$ed$ \\ \textbf{(EM)} \\ $\times 10^{-6}$ \end{tabular} \\
\hline
\begin{tabular}[c]{@{}l@{}} Rosenbluth \\ Extraction \cite{JohnsonMJ2013} \end{tabular} & 6.6196 & 6.6253 & 2.3954 & 2.1956 \\
\hline
\begin{tabular}[c]{@{}l@{}} Rosenbluth \\ Separation \cite{Bernauer000412214035042010} \end{tabular} & 832.92 & 841.93 & 225.39 & 238.93 \\
\hline
\begin{tabular}[c]{@{}l@{}} Friedrich- \\ Walcher \\ Model \cite{Bernauer000412214035042010} \end{tabular} & 737.26 & 744.84 & 197.24 & 204.42 \\
\hline
\begin{tabular}[c]{@{}l@{}} Polynomial \\ Model \cite{Bernauer000412214035042010} \end{tabular} & 738.73 & 746.31 & 204.99 & 212.17 \\
\hline
\begin{tabular}[c]{@{}l@{}} Spline \\ Model \cite{Bernauer000412214035042010} \end{tabular} & 739.24 & 746.80 & 204.82 & 211.86 \\
\end{tabular}
\end{ruledtabular}
\end{table}

For the Chi Test in Table \ref{ChiTestUD}, the $dcs_{ep}$ generated from different data sets were more in agreement with $edr_{*e*} dcs_{e*}$ than $edr_{*b*} dcs_{e*}$ since the deviation are much smaller in favor of the $dcs$ where the point particles assume effective quark masses. Again, it can also be seen that all the $dcs_{ep}$ are in more agreement with $edr_{u**} dcs_{eu}$ than with $edr_{d**} dcs_{ed}$ except for the Rosenbluth Extraction Data \cite{JohnsonMJ2013}. Expectedly, among the data sets from \cite{Bernauer000412214035042010}, the generated $dcs_{ep}$ from Rosenbluth Separation Data is the most favored by the Chi Test.

\begin{table}[htbp]
\caption{\label{ChiTestUD} Chi Test: The Chi Test between the $dcs_{ep}$ generated from the different data sets from their corresponding $dcs_{eu}$ and $dcs_{ed}$ with $edr$ where bare quark masses (BM) are used and, separately, for effective quark masses (EM).} 
\begin{ruledtabular}
\begin{tabular}{ccccc}
\begin{tabular}[c]{@{}l@{}} \textbf{Form Factor} \\ \textbf{Data Sets} \end{tabular} & \begin{tabular}[c]{@{}l@{}} $ep$-$eu$ \\ \textbf{(BM)} \\ $\times 10^{-6}$ \end{tabular} & \begin{tabular}[c]{@{}l@{}} $ep$-$ed$ \\ \textbf{(BM)} \\ $\times 10^{-6}$ \end{tabular} & \begin{tabular}[c]{@{}l@{}} $ep$-$eu$ \\ \textbf{(EM)} \\ $\times 10^{-6}$ \end{tabular} & \begin{tabular}[c]{@{}l@{}} $ep$-$ed$ \\ \textbf{(EM)} \\ $\times 10^{-6}$ \end{tabular} \\
\hline
\begin{tabular}[c]{@{}l@{}} Rosenbluth \\ Extraction \cite{JohnsonMJ2013} \end{tabular} & 8.9374 & 8.9499 & 4.2973 & 3.8496 \\
\hline
\begin{tabular}[c]{@{}l@{}} Rosenbluth \\ Separation \cite{Bernauer000412214035042010} \end{tabular} & 1647.3 & 1666.2 & 375.85 & 394.87 \\
\hline
\begin{tabular}[c]{@{}l@{}} Friedrich- \\ Walcher \\ Model \cite{Bernauer000412214035042010} \end{tabular} & 2565.1 & 2591.5 & 603.56 & 619.44 \\
\hline
\begin{tabular}[c]{@{}l@{}} Polynomial \\ Model \cite{Bernauer000412214035042010} \end{tabular} & 2557.6 & 2584.0 & 611.84 & 627.73 \\
\hline
\begin{tabular}[c]{@{}l@{}} Spline \\ Model \cite{Bernauer000412214035042010} \end{tabular} & 2558.0 & 2584.1 & 612.10 & 627.25 \\
\end{tabular}
\end{ruledtabular}
\end{table}

Considering Equation \ref{DCSEPScatteringPoint}, $edr$, weight contribution by quark flavor composition and, additionally, other criteria, four fitting models were formulated (see TABLE \ref{TheDCSModels}). The first is the Spin Bare Mass (SBM) which takes into account the respective contributions of $edr_{*bs}$ and $dcs_{e*}$. Second, is the Spin with other Criteria Bare Mass (SCBM) which is just the SBM but including the other considerations. The third is the Spin Effective Mass (SEM) which has lower amplitudes compared to the SBM and uses the effective quark masses. The fourth one, Spin with other Criteria Effective Mass (SCEM), is just the SEM but considering the same other criteria included in SCBM.

\begin{table}[htbp]
\caption{\label{TheDCSModels} The $dcs_{eq}$ Models: The four models include SBM, SCBM, SEM and SCEM and their forms.} 
\begin{ruledtabular}
\begin{tabular}{cc}
\textbf{Model} & \textbf{Form} \\
\hline
Spin Bare Mass (SBM) & \begin{tabular}[c]{@{}l@{}} $(2/3) 3.50 e^{-2.8 Q^2} dcs_{eu}$ \\ $+ (1/3) 14.0 e^{-2.8 Q^2} dcs_{ed}$ \end{tabular} \\
\hline
\begin{tabular}[c]{@{}l@{}} Spin with other Criteria\\ Bare Mass (SCBM) \end{tabular} & \begin{tabular}[c]{@{}l@{}} $(4/5) 3.50 e^{-2.8 Q^2} dcs_{eu}$ \\ $+ (1/5) 14.0 e^{-2.8 Q^2} dcs_{ed}$ \end{tabular} \\
\hline
Spin Effective Mass (SEM) & \begin{tabular}[c]{@{}l@{}} $(2/3) 2.40 e^{-2.8 Q^2} dcs_{eu}$ \\ $+ (1/3) 9.60 e^{-2.8 Q^2} dcs_{ed}$ \end{tabular} \\
\hline
\begin{tabular}[c]{@{}l@{}} Spin with other Criteria\\ Effective Mass (SCEM) \end{tabular} & \begin{tabular}[c]{@{}l@{}} $(4/5) 2.40 e^{-2.8 Q^2} dcs_{eu}$ \\ $+ (1/5) 9.60 e^{-2.8 Q^2} dcs_{ed}$ \end{tabular} \\
\end{tabular}
\end{ruledtabular}
\end{table}

The Ratio Test in TABLE \ref{RatioTestSC}, Absolute Difference in TABLE \ref{AbsoluteDifferenceSC} and Chi Test in TABLE \ref{ChiTestSC} show the comparisons of the data between the four fitting models and the corresponding generated $dcs_{ep}$ from different form factor data sets listed. The plots of the $dcs_{ep}$ from the Rosenbluth data sets with all the four models almost lie on the same space. It can be seen in TABLE \ref{RatioTestSC} that, in general, the $dcs_{ep}$'s are in agreement with SCBM for the ratio test. On the other hand, both $dcs_{ep}$'s from the Rosenbluth form factor data sets are in close agreement with SCEM.


\begin{table}[htbp]
\caption{\label{RatioTestSC} Ratio Test: The average ratio between the $dcs_{ep}$ of the different data sets to their corresponding $dcs_{eq}$ of the different models.} 
\begin{ruledtabular}
\begin{tabular}{ccccc}
\begin{tabular}[c]{@{}l@{}} \textbf{Form Factor} \\ \textbf{Data Sets} \end{tabular} & \textbf{SBM} & \textbf{SCBM} & \textbf{SEM} & \textbf{SCEM} \\
\hline
\begin{tabular}[c]{@{}l@{}} Rosenbluth \\ Extraction \cite{JohnsonMJ2013} \end{tabular} & 0.96967 & 0.96966 & 0.99933 & 0.99973 \\
\hline
\begin{tabular}[c]{@{}l@{}} Rosenbluth \\ Separation \cite{Bernauer000412214035042010} \end{tabular} & 1.0169 & 1.0168 & 0.98881 & 0.98909 \\
\hline
\begin{tabular}[c]{@{}l@{}} Friedrich-Walcher \\ Model \cite{Bernauer000412214035042010} \end{tabular} & 1.0505 & 1.0505 & 1.1510 & 1.1514 \\
\hline
\begin{tabular}[c]{@{}l@{}} Polynomial \\ Model \cite{Bernauer000412214035042010} \end{tabular} & 1.2054 & 1.2053 & 1.3443 & 1.3448 \\
\hline
\begin{tabular}[c]{@{}l@{}} Spline \\ Model \cite{Bernauer000412214035042010} \end{tabular} & 1.2135 & 1.2135 & 1.3546 & 1.3550 \\
\end{tabular}
\end{ruledtabular}
\end{table}

For the comparison using Absolute Difference in TABLE \ref{AbsoluteDifferenceSC}, SCBM is favored over SBM by all the generated $dcs_{ep}$ from different form factor data sets. In general, SCEM is also favored by the generated $dcs_{ep}$ except those generated from Rosenbluth Extraction Data from \cite{JohnsonMJ2013} and this could be due to the experimental parameters in considerations. With the numbers given in this table, the SCEM is most favored since its corresponding average absolute difference is smaller compared to SCBM; both fitting models feature the other additional criteria.

\begin{table}[htbp]
\caption{\label{AbsoluteDifferenceSC} Absolute Difference: The average absolute difference between the $dcs_{ep}$ of the different data sets and their corresponding $dcs_{eq}$ of the different models|Spin Bare Mass (SBM), Spin with other Criteria Bare Mass (SCBM), Spin Effective Mass (SEM) and Spin with other Criteria Effective Mass (SCEM).} 
\begin{ruledtabular}
\begin{tabular}{ccccc}
\begin{tabular}[c]{@{}l@{}} \textbf{Form Factor} \\ \textbf{Data Sets} \end{tabular} & \begin{tabular}[c]{@{}l@{}} \textbf{SBM} \\ $\times 10^{-6}$ \end{tabular} & \begin{tabular}[c]{@{}l@{}} \textbf{SCBM} \\ $\times 10^{-6}$ \end{tabular} & \begin{tabular}[c]{@{}l@{}} \textbf{SEM} \\ $\times 10^{-6}$ \end{tabular} & \begin{tabular}[c]{@{}l@{}} \textbf{SCEM} \\ $\times 10^{-6}$ \end{tabular} \\
\hline
\begin{tabular}[c]{@{}l@{}} Rosenbluth \\ Extraction \cite{JohnsonMJ2013} \end{tabular} & 6.6215 & 6.6207 & 2.3035 & 2.3403 \\
\hline
\begin{tabular}[c]{@{}l@{}} Rosenbluth \\ Separation \cite{Bernauer000412214035042010} \end{tabular} & 835.92 & 834.72 & 229.90 & 228.09 \\
\hline
\begin{tabular}[c]{@{}l@{}} Friedrich- \\ Walcher \\ Model \cite{Bernauer000412214035042010} \end{tabular} & 739.79 & 738.78 & 199.64 & 198.68 \\
\hline
\begin{tabular}[c]{@{}l@{}} Polynomial \\ Model \cite{Bernauer000412214035042010} \end{tabular} & 741.26 & 740.25 & 207.38 & 206.42 \\
\hline
\begin{tabular}[c]{@{}l@{}} Spline \\ Model \cite{Bernauer000412214035042010} \end{tabular} & 741.76 & 740.76 & 207.16 & 206.22 \\
\end{tabular}
\end{ruledtabular}
\end{table}

\begin{figure} [htbp]
\centering
\includegraphics[width = 10.25cm, height = 8.25cm, keepaspectratio]{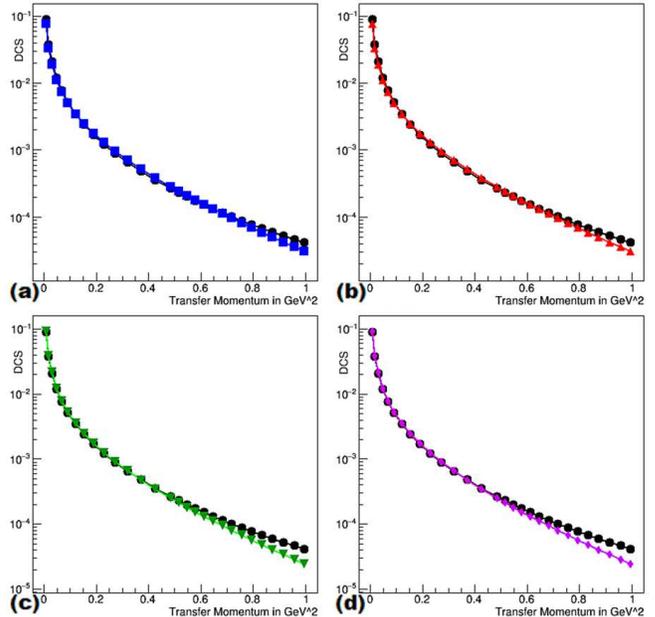}
\caption{The plot shows the generated $dcs_{ep}$ (black $\bullet$) using the form factors from Friedrich-Walcher model \cite{Bernauer000412214035042010} versus $Q^2$. They were compared to \textbf{(a)} $dcs_{SBM}$ ($\blacksquare$), \textbf{(b)} $dcs_{SCBM}$ ($\blacktriangle$), \textbf{(c)} $dcs_{SEM}$ ($\blacktriangledown$) and \textbf{(d)} $dcs_{SCEM}$ ($\blacklozenge$), showing a pronounced agreement.}
\label{Figure344DataPlotsFriedrichWalcher}
\end{figure}

\begin{figure} [htbp]
\centering
\includegraphics[width = 10.25cm, height = 8.25cm, keepaspectratio]{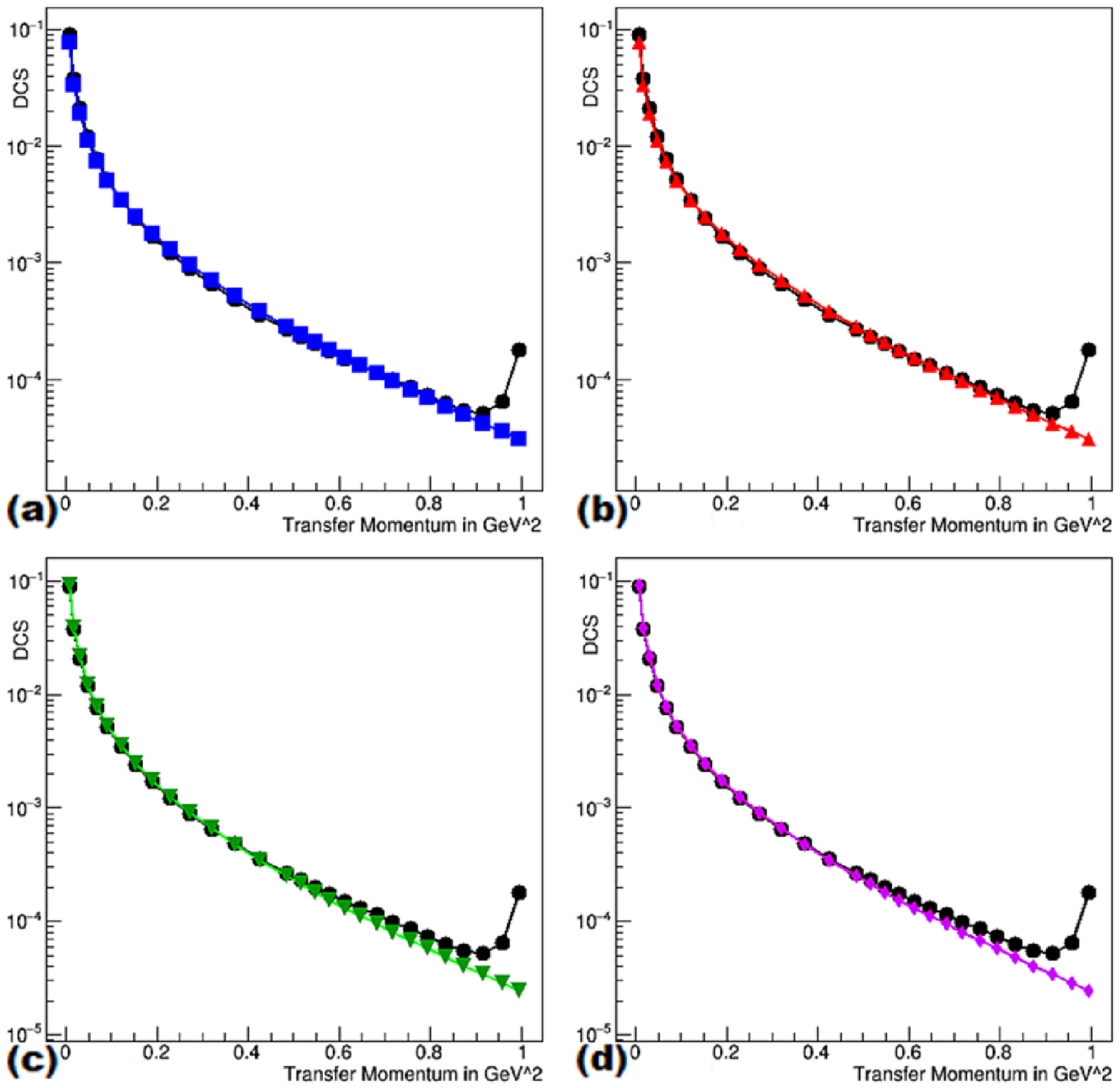}
\caption{The plot shows the generated $dcs_{ep}$ (black $\bullet$) using the form factors from Spline model \cite{Bernauer000412214035042010} versus $Q^2$. They were compared to \textbf{(a)} $dcs_{SBM}$ ($\blacksquare$), \textbf{(b)} $dcs_{SCBM}$ ($\blacktriangle$), \textbf{(c)} $dcs_{SEM}$ ($\blacktriangledown$) and \textbf{(d)} $dcs_{SCEM}$ ($\blacklozenge$), showing good agreement except with the two points at the tail.}
\label{Figure345DataPlotsBernauerSpline}
\end{figure}

\begin{figure} [htbp]
\centering
\includegraphics[width = 10.25cm, height = 8.25cm, keepaspectratio]{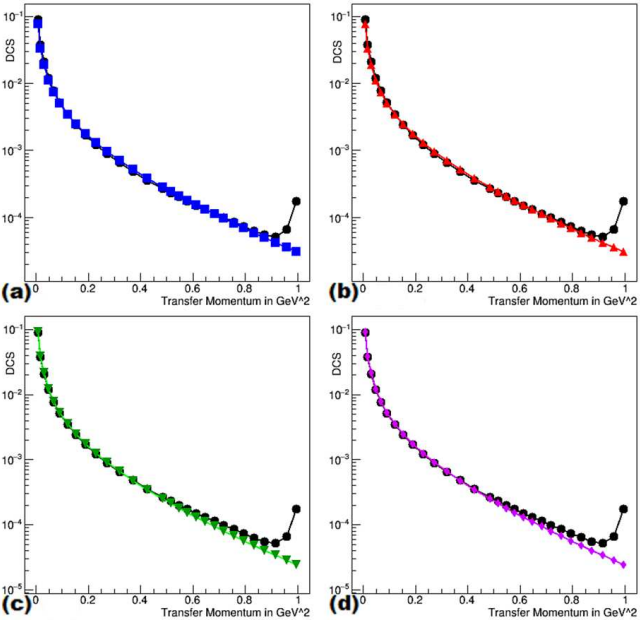}
\caption{The plot shows the generated $dcs_{ep}$ (black $\bullet$) using the form factors from Polynomial model \cite{Bernauer000412214035042010} versus $Q^2$. They were compared to \textbf{(a)} $dcs_{SBM}$ ($\blacksquare$), \textbf{(b)} $dcs_{SCBM}$ ($\blacktriangle$), \textbf{(c)} $dcs_{SEM}$ ($\blacktriangledown$) and \textbf{(d)} $dcs_{SCEM}$ ($\blacklozenge$), showing good agreement except with the two points at the tail.}
\label{Figure346DataPlotsBernauerPolynomial}
\end{figure}

The plots in FIG. \ref{Figure344DataPlotsFriedrichWalcher}, FIG. \ref{Figure345DataPlotsBernauerSpline} and FIG. \ref{Figure346DataPlotsBernauerPolynomial} show the $dcs_{ep}$ of form factors derived from Friedrich-Walcher, Spline and Polynomial models, respectively, together with the formulated fitting models. From these three data sets, it is the generated $dcs_{ep}$ from the Friedrich-Walcher form factors that has the smallest average absolute difference. Looking at FIG. \ref{Figure345DataPlotsBernauerSpline} and FIG. \ref{Figure346DataPlotsBernauerPolynomial}, it can be seen that the last two data points from the data sets diverge way-off from the models and this could be attributed by the limitations of the experimental set-up and the fitting parameters when the form factors were derived. It is also only up to this region that the formulated fitting models are expected to be valid.

For the comparison using Chi Test in TABLE \ref{ChiTestSC}, it is expected that the Rosenbluth form factor data sets are favorable to all four models since the Chi Test values are smaller, compared to the other data sets; but more specially to SCBM and SCEM. With general considerations, it is the SCEM that is the most favored model for this comparison test with SEM coming next.

\begin{table}[htbp]
\caption{\label{ChiTestSC} Chi Test: The Chi Test between the $dcs_{ep}$ of the different data sets and their corresponding $dcs_{eq}$ of the different models|Spin Bare Mass (SBM), Spin with other Criteria Bare Mass (SCBM), Spin Effective Mass (SEM) and Spin with other Criteria Effective Mass (SCEM).} 
\begin{ruledtabular}
\begin{tabular}{ccccc}
\begin{tabular}[c]{@{}l@{}} \textbf{Form Factor} \\ \textbf{Data Sets} \end{tabular} & \begin{tabular}[c]{@{}l@{}} \textbf{SBM} \\ $\times 10^{-6}$ \end{tabular} & \begin{tabular}[c]{@{}l@{}} \textbf{SCBM} \\ $\times 10^{-6}$ \end{tabular} & \begin{tabular}[c]{@{}l@{}} \textbf{SEM} \\ $\times 10^{-6}$ \end{tabular} & \begin{tabular}[c]{@{}l@{}} \textbf{SCEM} \\ $\times 10^{-6}$ \end{tabular} \\
\hline
\begin{tabular}[c]{@{}l@{}} Rosenbluth \\ Extraction \cite{JohnsonMJ2013} \end{tabular} & 8.9416 & 8.9399 & 4.1442 & 4.2050 \\
\hline
\begin{tabular}[c]{@{}l@{}} Rosenbluth \\ Separation \cite{Bernauer000412214035042010} \end{tabular} & 1653.6 & 1651.1 & 382.19 & 379.65 \\
\hline
\begin{tabular}[c]{@{}l@{}} Friedrich- \\ Walcher \\ Model \cite{Bernauer000412214035042010} \end{tabular} & 2573.9 & 2570.4 & 608.85 & 606.73 \\
\hline
\begin{tabular}[c]{@{}l@{}} Polynomial \\ Model \cite{Bernauer000412214035042010} \end{tabular} & 2566.4 & 2562.9 & 617.13 & 615.01 \\
\hline
\begin{tabular}[c]{@{}l@{}} Spline \\ Model \cite{Bernauer000412214035042010} \end{tabular} & 2566.7 & 2563.2 & 617.14 & 615.13 \\
\end{tabular}
\end{ruledtabular}
\end{table}

\section{Conclusions and Recommendations}

Several experimental data, such as those coming from A1-Collaboration and JLab, have measured the proton electromagnetic form factors with precision and accuracy for relativistic systems through elastic scatterings. These measurements, specially for $Q^2 < 1GeV^2$, are important since they give the electric and magnetic form factors that determine the distribution of charge and magnetization of the proton or its charge and magnetic (rms) radii.

The $dcs_{ep}$ generated from different sets of form factor data were compared to raw $dcs_{eq}$ where $q$ is a point particle assigned with bare and effective masses of $u$ and $d$ quarks.
The $edr$'s were determined from this comparison and are listed in TABLE \ref{EnergyDecayingRatio}.
The $edr$ that suit best the generated data corresponds to the one derived from Rosenbluth Separation Data in \cite{Bernauer000412214035042010}.
The amplitude of $edr_{d**}$ is greater than that of $edr_{u**}$ and this is due to their differences in charge and, eventually, in mass as transfer momentum increments.
It is recommended that this will be delved more; specially, on the behavior of the ratio $edr_{d**}/edr_{u**}$. 
Also, $edr_{*e*} < edr_{*b*}$ and this could be due to the dominance of the constituent or effective mass at the range of transfer momentum studied.
Aside from that, it is quite logical that the point particle with (smaller) bare mass would need a thicker cloud to compensate for its mass compared to the one with (bigger) effective mass.
The decay rate in the $edr$ is constant, however, this could change depending on the number of data points considered in the formulation of the fitting model or if different form factor data sets are used, in addition to the speculation that this variation could also be due to the differences in the set-up and parameters considered in the experiments; as can be seen, $edr_{**s} > edr_{**e}$.
By averaging the $dcs$ of 2000 events, each taken with different and randomly selected scattering angles from $0^o$ to $180^o$, the recoil factor can be treated as a constant.
Moreover, it was necessary to modify the recoil factor of the $eq$-scattering by using the proton mass to shift the distribution of $dcs_{eq}$ closer to $dcs_{ep}$.
And, this materializes the proton as a parameter to the fitting model.
The existence of the $edr$ and the modification of the recoil factors, foremost, are acting as the cloud layers that are supposed to cover the point particle $q$ at low energy.
Furthermore, TABLE \ref{RatioTestUD}, TABLE \ref{AbsoluteDifferenceUD} and TABLE \ref{ChiTestUD} imply with generality that the generated $dcs_{ep}$ favors $edr_{u**}dcs_{eu}$ over $edr_{d**}dcs_{ed}$.

Four models were formulated (see TABLE \ref{TheDCSModels}) considering the assignment of bare and effective quark masses|SBM and SEM consider the $edr$ and contributions based on the quark flavor composition of the proton while SCBM and SCEM incorporate other considerations, albeit arbitrarily, such as overlapping of the electron wave functions, spin-spin interactions, and color interactions. For the Ratio Test, SCBM is the most favored model while SCEM is favored by both the Absolute Difference and Chi Test. With SCEM and SEM having favorable comparative numbers imply that at this range of transfer momenta, the said models are consistent on the point particles being likely to assume effective masses rather than bare masses. It should be noted that the fitting models are not meant to prove the quark composition of the proton but, rather, show that previous and known results of $eq$-scattering with modifications can be used to create models for $ep$-scattering at low transfer momenta.

Although the additional arbitrary considerations has an effect to the elastic $ep$-scattering, it is assumed to be really very small in magnitude for $Q^2 < 1GeV^2$ but its existence in the models have been very helpful in optimizing the comparison tests as manifested by both the SCBM and SCEM models. It is recommended, for example, to re-assess the geometrical arrangement preferences of the quarks and gluons and the configuration counting in order to have a more optimized fitting model. Variations in the results are also expected by considering more number of events and thus involving more scattering angles. The cloud covering is also affected by the overlapping of the electron's initial and final wave functions and the overall spin-spin interactions between the electron and proton. These effects will be investigated more.

\section*{Acknowledgement}

The Mindanao State University - Iligan Institute of Technology (MSU-IIT) and its Department of Physics and the Premier Research Institute for Science and Mathematics (PRISM) of Iligan City, Philippines; Research Center for Theoretical Physics (RCTP) of Jagna, Philippines; and Centro de Investigacion en Computacion - Instituto Politecnico Nacional (CIC-IPN) of CDMX, Mexico are acknowledged for their conducive venues in making this research possible. Gratitude is extended to the Department of Science and Technology (DOST) of the Philippines and MSU-IIT for their financial support. The inspiration and encouragements from Prof. Christopher Bernido, Prof. Maria Victoria Bernido, Prof. Ludwig Streit, and Prof. Roland Winkler are highly appreciated.


\end{document}